\documentclass[usenatbib]{mn2e}

\bibpunct{(}{)}{;}{a}{}{,}
\usepackage{epsfig}
\renewcommand{\vec}[1]{\mbox{\boldmath$#1$}}
\renewcommand{\d}[2]{\frac{\partial #1}{\partial #2}}
\newcommand{\dt}[1]{\d{#1}{t}}

\newcommand{\Div}{\vec \nabla \! \cdot \!}

\newcommand{\Grad}{\vec \nabla}
\newcommand{\otprod}{\! \otimes \!}           
\newcommand{\Nabla}{\vec \nabla} 
\newcommand{\cross}{\times}      

\renewcommand{\r}{r}             
\newcommand{\R}{R}          
\newcommand{\Z}{z}               

\newcommand{\grav}{\Phi}         
\newcommand{\Amag}{A}            
\newcommand{\kflux}{\Psi_\Amag}  
\newcommand{\entrop}{Q}          %
\newcommand{\pol}{_{\rm p}}      
\newcommand{\eu}{\vec {\hat e}}  
\newcommand{\num}{{\rm N}}
\newcommand{\vap}{v_{\rm Ap}}     
\newcommand{\vf}{v_{\rm f}}      
\newcommand{\vs}{v_{\rm s}}      

\newcommand{\Ref}[1]{(\ref{#1})}

\newcommand{\VTST}{2000MNRAS.318..417V}
\newcommand{\BP}{1982MNRAS.199..883B}
\newcommand{\VT}{1998MNRAS.298..777V}
\newcommand{\KLBi}{1999ApJ...526..631K}
\newcommand{\KLBii}{2003ApJ...595..631K}

\begin{document}

\title{Jet simulations extending radially self-similar MHD models}

\author[J. Gracia, N. Vlahakis and K. Tsinganos] {
  J. Gracia\thanks{E-mail: jgracia@phys.uoa.gr;
  vlahakis@phys.uoa.gr; tsingan@phys.uoa.gr}, 
  N. Vlahakis\footnotemark[1]
  and K. Tsinganos\footnotemark[1]\\
  IASA and Section of Astrophysics, Astronomy and Mechanics,
  Department of Physics, University of Athens,\\ 
  Panepistimiopolis, GR-157\,84 Zografos, Athens, Greece
}

\date{Accepted 2005. Received 2005; in original form}

\pagerange{\pageref{firstpage}--\pageref{lastpage}} \pubyear{2005}

\maketitle

\label{firstpage}

\begin{abstract}
We perform a numerical simulation of magnetohydrodynamic radially
self-similar jets, whose prototype is the Blandford \& Payne
analytical example.  The reached final steady state is valid close to
the rotation axis and also at large distances above the disk where the
classical analytical model fails to provide physically acceptable
solutions.  The outflow starts with a sub-slow magnetosonic speed which
subsequently crosses all relevant MHD critical points and
corresponding magnetosonic separatrix surfaces.  The characteristics
are plotted together with the Mach cones
and the super-fast magnetosonic outflow satisfies MHD causality.  The
final solution remains close enough to the analytical one which is
thus shown to be topologically stable and robust for various boundary
conditions.
\end{abstract}


\section{Introduction}
Cosmic jets are ubiquitous being quite often associated with
newborn stars, X-ray binaries, AGN and GRBs. In all such cases jets and
disks seem to be interrelated.  Not only jets need disks in order to
provide them with the ejected plasma and magnetic fields, but also
disks need jets in order that the accreted plasma gets rid of its
excess angular momentum to accrete and observationally there has been
already accumulated enough evidence for such a correlation.  For
example, in star forming regions 
an apparent correlation is found between accretion diagnostics and
outflow signatures \citep{1995ApJ...452..736H}. Hence, our current
understanding is that jets are fed by the material of an accretion
disk surrounding the central object. Energetically, most jets are also
believed to be powered by the gravitational energy released in the
accretion process.  Thus, in the case of jets associated with YSOs it
is well known that the observed mechanical luminosity in the bipolar
outflows is typically a factor $\sim 10^2$ higher than the total
radiant luminosity of the embedded central object
\citep{1985ARA&A..23..267L}, a fact that seems to rule out radiative
acceleration of those jets. Furthermore, the kinetic luminosity of the
outflow seems to be a large fraction of the rate at which energy is
released by accretion. With such a high ejection efficiency it is
natural to assume that jets are driven magnetically from an accretion
disk; the magnetic model of a disk-wind seems to explain
simultaneously acceleration, collimation as well as the observed high
jet speeds \citep{2000prpl.conf..759K}.

The first analytical model of a magnetised disk-wind which
demonstrated that a cold plasma can be launched magneto-centrifugally
from a Keplerian disk \citep{\BP}, has been shown to belong to the
wide family of radially self-similar solutions of the full MHD
equations \citep{\VT}. This model however has basically two serious
limitations.  {\it First}, the outflow speed at large distances does
not crosses the corresponding limiting characteristic, with the result
that the terminal wind solution is not causally disconnected from the
disk. This shortcoming however of the original \citet{\BP} model has
been remedied in \citet{\VTST} where it has been shown that a terminal
wind solution can be constructed which is causally disconnected from
the disk and hence any perturbation downstream of the super-fast
transition cannot affect the whole structure of the steady outflow.
The {\it second} limitation of most radially self-similar analytic
models is that they must be cutoff at small cylindrical radii and also
at some vertical height; essentially this has to do with the
singularity these models posses at the system axis wherein the
electric current diverges.  As a result, the solutions terminate at
some height above the disk because of the strong Lorentz force
pinching the outflow and magnetic field towards the axis. The purpose
of this article is to address this question, namely if radially
self-similar solutions can be extended numerically at small
cylindrical distances from the axis and also at large heights above
the disk. 

At the same time numerical studies have addressed the problem of a
magnetised disk-wind to an underlying disk. An incomplete selection of
these are \citet{\KLBi, \KLBii}, \citet{2004ApJ...601...90C},
\citet{1995ApJ...439L..39U}, \citet{2003ApJ...582..292O},
\citet{2004ApJ...617..123N}. The relation and similarities of these
numerical studies to the present work will be presented in
the last section.

This paper is structured as follows. In the following section we
briefly summarise the basic analytical expressions of a radially
self-similar jet model. In \S 3 we describe the initial and boundary
conditions employed in this work. Our results are presented in
\S 4 with emphasis on the critical MHD surfaces and characteristics,
the MHD integrals of motion and the acceleration and collimation of
the outflow. Section 5 is devoted to a discussion of the robustness of
the numerical solution and comparison to other analytical and
numerical studies.

\section{The analytical model}
We will use two different coordinate systems, cylindrical $(\Z, \R,
\phi)$, and spherical $(\r, \theta, \phi)$ coordinates, and express
all equations using the international system of units (SI).

The non-relativistic ideal MHD equations are 
\begin{equation} \label{eq1}
  \dt{\rho} = - \Div (\rho \vec V),
\end{equation}
\begin{equation}
  \dt{\rho \vec V} = - \Div (\rho \vec V \otprod \vec V) - \Grad P 
  - \rho \Grad \grav 
  + \frac{(\Nabla \cross \vec B) \cross \vec B}{\mu_0},
\end{equation}
\begin{equation}
  \dt{\vec B} = \Nabla \cross (\vec V \cross \vec B),
\end{equation}
\begin{equation}\label{eq:energy}
  \dt{e} = - \Div (e\vec V) - P \Div \vec V,
\end{equation}
where $\vec V$ is the flow velocity, $\vec B$ the magnetic field,
$(\rho, P)$ the gas density and pressure, and
$\grav = -{\cal GM}/\r$ the gravitational potential.
The internal energy density $e$ is related to the pressure $P$ by 
\begin{equation}\label{eq:Pideal}
  P = (\gamma - 1) e,
\end{equation}
where $\gamma$ is the effective polytropic index.

Assuming steady state and axisymmetry, there are at least two families
of exact solutions available, namely the meridionally and radially
self-similar solutions \citep{\VT}. For example, a special class of
solutions of the first type has been presented in
\citet{1994A&A...287..893S} and of the latter type -- actually an
extension of the classical \citet{\BP} model -- was studied in
\citet{\VTST}.

Under the assumptions of axisymmetry, steady-state, and radial
self-similarity, the physical quantities are given by the following
expressions
\begin{equation}
  \frac{\rho}{\rho_0} = \alpha^{x-3/2} \frac{1}{M^2},
\end{equation}
\begin{equation}
  \frac{P}{P_0} = \alpha^{x-2} \frac{1}{M^{2\gamma}},
\end{equation}
\begin{equation} \label{eq:Bp}
 \frac{\vec B\pol}{B_0} = - \alpha^{\frac{x}{2}-1} \frac{1}{G^2} 
       \frac{\sin \theta}{\cos(\psi + \theta)} \; 
\left(\eu_\R \cos \psi + \eu_\Z \sin \psi\right),
\end{equation}
\begin{equation}
  \frac{\vec V\pol}{V_0} = -\alpha^{-1/4} \frac{M^2}{G^2} 
       \frac{\sin \theta}{\cos(\psi + \theta)} \; 
       \left(\eu_\R \cos \psi + \eu_\Z \sin \psi\right),
\end{equation}
\begin{equation}
  \frac{B_\phi}{B_0} = -\lambda \alpha^{\frac{x}{2}-1} 
  \frac{1-G^2}{G (1-M^2)},
\end{equation}
\begin{equation}
  \frac{V_\phi}{V_0} = \lambda \alpha^{-1/4} \frac{G^2 - M^2}{G(1-M^2)},
\end{equation}
where $G = G(\theta)$, $M = M(\theta)$ and $\psi = \psi(\theta)$ are
functions of $\theta$ only, and
\begin{equation}\label{eq:alpha}
  \alpha = \frac{\r^2}{\R_0^2} \frac{\sin^2\theta}{G^2} 
          = \frac{\R^2}{\R_0^2 G^2}.
\end{equation}
The reference length $\R_0$, magnetic field $B_0$ and mass of the
central object $\cal M$ can be freely chosen, while the remaining
normalisation constants $(V_0, \rho_0, P_0)$ are given by the
following relations
\begin{equation}
  \kappa V_0 = \sqrt{\frac{\cal GM}{\R_0}}, 
  \qquad \frac{B_0}{\sqrt{\mu_0\rho_0}} = V_0, \qquad P_0 = \mu \frac{B_0^2}{2\mu_0}, \qquad.
\end{equation}

Note, that the pressure and density are related via a polytropic
equation $P = \entrop(\alpha) \, \rho^\gamma$, with the entropy function
$\entrop = \entrop(\alpha)$ constant along a flowline, but different
from one flowline to the other. This relation is the general steady
solution of equation (\ref{eq:energy}) for an equation of state given
by (\ref{eq:Pideal}).  The system of MHD equations is then reduced to
three first order ordinary differential equations (ODEs) with respect
to the functions $G(\theta)$, $M(\theta)$ and $\psi(\theta)$.  A
particular solution is given in terms of the set of formal solution
parameters $(x, \lambda^2, \mu, \kappa, \gamma)$ and a prescription
for the solution functions $(G, M, \psi)$, which are calculated
numerically by solving the ODEs.  The final remaining difficulty is to
ensure that the flow crosses the three singular MHD surfaces, where
the appropriate regularity conditions need to be applied.  Such a
solution satisfying this rather important constraint has been
presented in \citet{\VTST}.

By virtue of the radial self-similarity assumption this class of
solution breaks down in general at the symmetry axis, where $1/\alpha
\propto \R^{-2}$ diverges. Another deficiency of these self-similar
solutions is that they are terminated at a finite height above the
disk. A third point to note is that radially self-similar flows do not
have an intrinsic scale.

\section{The numerical model} 
The aim of the numerical work, is to extent the analytical solution in
two respects. First, into the domain where the self-similar assumption
breaks down. That means, extending the solution towards and including
the axis, while testing if the self-similar solution is a good
approximation away from the axis; determining where this self-similar
regime is located, and describing the differences between the
self-similar solution and the numerical solution. In addition, the
solution will be extended to large distances, where the self-similar
description gives terminated solutions.  Second, coupling the
self-similar solution to boundary conditions, which are {\em not}
self-similar but have a typical scale, like, e.g., the accretion disk
size, and understanding the resulting MHD flow in terms of the
analytical solution, if possible. This will allow to apply the
analytical models to actual astrophysical systems. In the present work
we fulfil the first aim; we leave the treatment of the second task to
a future publication. 

The numerical work has been done with the grid-based, time-dependent
MHD code {\em NIRVANA} \citep{Ziegler}. This code explicitly solves
the system of equations \Ref{eq1}-\Ref{eq:Pideal}. Note, that the
energy equation \Ref{eq:energy} is solved explicitly instead of using
a polytropic relation between pressure and density.
There are a number of differences between the analytical model and the
numerical model. The first and most obvious arises from the applied
modifications discussed in the following paragraphs. They are
necessary to make the numerical model physically consistent on the
axis, and reflects the fact, that numerical models do always have a
finite spatial resolution, while analytical models are continuous. We
shall briefly discuss some of the most obvious implications.

It is well known that along each fieldline a number of physical
quantities are conserved for steady-state, axisymmetric, ideal MHD
flows, such as the specific entropy or the total specific angular
momentum. In other words, the surfaces of constant values of these
conserved {\em integrals of motion} and the surfaces of constant
magnetic flux $\Amag$ coincide. Due to the modifications made to the
magnetic and velocity fields this is no longer true for the numerical
model. This alone makes the initial conditions non-steady. More
important is though, that on the boundary the relation of magnetic
flux to any integral quantity changes with respect to the analytical
model. So, the final steady state of the numerical simulation will be
different from the underlying analytical model.


The numerical simulations are carried out on a cylindrical grid
of size $\R = \epsilon \R_0 - 50 \R_0$ and $\Z = 6 \R_0 - 100 \R_0$,
where $\epsilon \approx 0$ is much smaller than the typical cell
size. We did not consider the region near the equator at $\Z = 0$,
but instead chose a finite lower $\Z$ boundary to avoid very
small numerical time-steps near the equator.

\subsection{Initial conditions}
A particular solution to the previously discussed self-similar
analytical model consisting of a set of solution parameters and the
tabulated solution functions $G(\theta), M(\theta), \psi(\theta)$ is
adopted as the basis for the initial conditions. These functions are
not available for very small values of $\theta<0.025 \, {\rm rad}$,
i.e. very close to the axis.  Since the numerical model
includes the symmetry axis, some modifications had to be done. These
modifications are applied in the order in which they are described in
the following paragraphs. Subscripts $_\num$ or superscripts $^\num$
indicate numerical relations differing from the analytical ones.

The tabulated solution functions $(G, M, \psi)$ have to be
extrapolated for small angles near the axis. It turned out, that
straightforward extrapolation, either first or second order, does not
yield suitable quantities. Instead, we exploited the fact, that the
solution functions are symmetric around the axis, i.e. $G(-\theta) =
G(\theta)$ and {\em interpolate} small angles with an Akima-type
interpolator from the GNU Scientific Library. This interpolation,
rather than extrapolation, yields smooth results in all
quantities. The same interpolator is used for other non-sampled values
of $\theta$. 

All physical quantities depend on the function $1/\alpha
\propto \R^{-2}$, which diverges on the axis. Even if this is not
necessary, due to the staggered-grid that our code uses, one might
avoid problems near the axis by smoothing $\alpha$ and replacing it
with 
\begin{equation}
  \alpha_\num = \frac{(\R + \R_0)^2}{\R_0^2 G^2}.  
\end{equation}
We checked that such modifications do not change the results or
conclusions in this paper.

The next modification affects the magnetic field only.  While the
analytical solution is locally $\Div B$-free everywhere in its domain
of availability, the asymptotic continuation towards the axis is $\Div
B$-free only at the expense of diverging magnetic field strength on
the axis. Further, to guarantee the solenoidal character of the magnetic
field, the function $G(\theta)$ must approach its asymptotic value on
the axis in a very special way given by solving $\Div \vec B = \Div
\vec B\pol = 0$ near and including the axis. This is something, which
our extrapolation cannot take into account. A diverging magnetic field
strength is also not desirable from a physical point of
view. 

Therefore, the numerical model adopts a modified magnetic field
structure. However, this affects only the poloidal components; the
azimuthal component $B_\phi$ remains unchanged since the model is
axisymmetric.  Instead of using the relation \Ref{eq:Bp} for the
poloidal magnetic field components directly, we calculate the magnetic
flux function $\Amag$ as given by \citet{\VTST} from the extrapolated
solution functions and impose the desired boundary conditions on the
axis, i.e.
\begin{equation} \label{eq:Anum}
  \Amag_\num = \frac{B_0 \R_0^2}{x} \alpha^{x/2}, 
  \qquad B_\R^\num(\theta=0) = 0.
\end{equation}
The vertical magnetic field component $B_\Z^\num$ is then calculated
from the definition of the magnetic flux function
\begin{equation}\label{eq:Amag}
  \vec B\pol = \frac{\Grad \Amag \cross \eu_\phi}{\R},
\end{equation}
and the other, i.e. $B_\R^\num$, from the $\Div B$-free constraint, to
impose this constraint also numerically down to machine precision
level. This procedure yields a magnetic field, which is very close to
the analytical one, but physically consistent on the axis and carries
over easily to coordinate systems other than cylindrical where it can
be expected to give similar results.

The last modification concerns the velocity field. One of the
properties of steady, axisymmetric, ideal MHD flows, is that the
poloidal components of the velocity and magnetic field are parallel or
anti-parallel, i.e.$\vec V\pol || \vec B\pol$. Since the magnetic field
has been modified, a similar modification has to be applied to the
velocity field. Strictly speaking, this is not necessary for the whole
computational domain, but
has to be enforced at least on the boundary. Physically, a
non-alignment of the two fields results in an inductive electric field
$\vec E_\phi = - \vec V\pol \cross \vec B\pol$ and could make the
magnetic field non-steady. In practice we calculate the velocity field
of the numerical model by making the poloidal velocity parallel to the
poloidal numerical magnetic field, i.e.$\vec V\pol^\num || \vec
B\pol^\num$, and keeping the analytical 
value of the vertical velocity component, $V_\Z^\num = V_\Z$.  This
approach keeps the mass-to-magnetic-flux ratio close to the analytical
value.

\subsection{Boundary conditions}
The numerical initial conditions on the boundaries of the
computational domain are taken as fixed boundary conditions and are,
in principle, not modified during the simulation. Only if the velocity
component perpendicular to the boundary points out of the
computational box, then the physical quantities immediately inside the
computational domain, are copied to the boundary allowing the flow to
propagate outward. This is practically always the case at the upper
$\Z$ and the outer $\R$ boundary, while the lower $\Z$ boundary
usually enforces the analytical solution to propagate into the
computational domain, but still allows backflows to be absorbed. The
lower $\R$ boundary enforces the proper symmetry conditions on the
rotation axis.  The boundary conditions for the magnetic field impose
only the components non-perpendicular to the boundary from the initial
model. The perpendicular component is calculated to guarantee $\Div
\vec B = 0$.

Numerical values for seven physical quantities have to be provided on
the boundary, i.e. density, pressure, three velocity components, the
azimuthal and one poloidal magnetic field component, while the other
is given from the $\Div \vec B$-free constraint. Strictly speaking,
the boundary conditions are over-specified in a mathematical sense, if
all seven values are chosen independently. The number of boundary
conditions should be reduced by one for each critical surface that is
crossed downstream, since the corresponding waves cannot propagate
upstream from those critical surfaces. For example, in the sub-slow
regime far away from the axis, only four conditions should be
enforced, while near the axis, where the flow is super-fast, all seven
conditions must be imposed. The remaining quantities should be
extrapolated in a suitable way from the computational domain, to allow
the boundaries to adjust to the regularity conditions at the critical
surfaces.

In the present work, we still choose to impose all seven quantities
as described in the previous paragraphs. This allows us to relax the
initial conditions into a numerical solution which overcomes the
shortcomings of the analytical model, while sticking to it as close
as possible.  In the analytical solution the azimuthal vector
components diverge on the axis, which is physically inconsistent,
while the numerical solution enforces the correct vanishing of these,
due to the symmetry conditions. This leads to a steep gradient in
$B_\phi$ and $V_\phi$ near the axis, and causes numerical problems if
the correct number of boundary conditions is imposed. While smoothing
of the azimuthal vector components solves these numerical problems, it
introduces additional degrees of freedom and differences between the
analytical and numerical model, which we want to prevent. On
the other hand, no artifacts from this over-specification are obvious
at the moment.

\section{Results}
\begin{figure}
  \resizebox{\hsize}{!}{\includegraphics{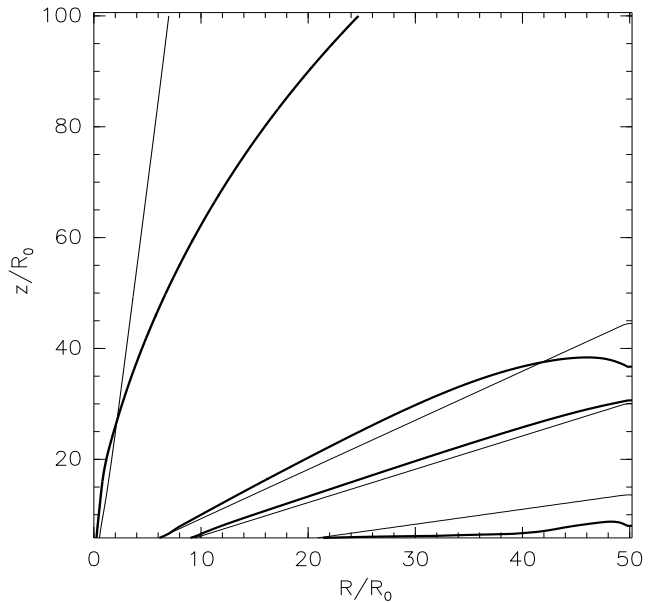}}
  \caption{ \label{fig:MHD_surfaces} Critical surfaces in the poloidal
  plane, with thin lines corresponding to the initial model and heavy
  lines to the final model. The individual lines represent from
  bottom to top the slow magnetosonic critical ($V\pol = \vs$),
  Alfv\'enic critical ($V\pol = \vap$), the fast magnetosonic critical
  surface ($V\pol = \vf$) and the limiting characteristic surface or
  FMSS (see text), respectively.}
\end{figure}
\begin{figure}
  \resizebox{\hsize}{!}{\includegraphics{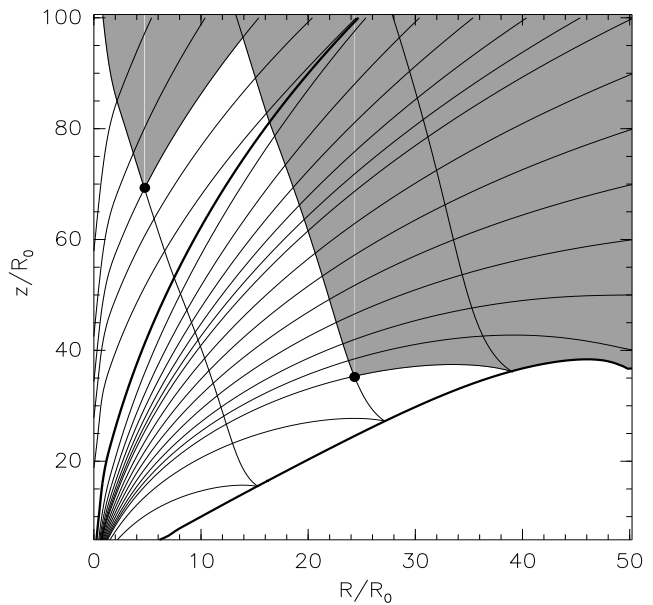}}
  \caption{ \label{fig:characteristics} Topology of the
  characteristics and Mach cones for the final state. The thin lines
  indicate the two distinct families of characteristics. The lower heavy
  line indicates the classical fast surface, the upper heavy line the
  modified-fast surface or limiting characteristic.  Two families of
  characteristics originate from the classical fast surface. For each
  point in space they spawn the Mach cone, i.e. the area which is
  causally connected to the origin.  The Mach cones originating in the
  two points indicated by black circles are shown as shaded regions.
  One Mach cone originates above the modified-fast surface, the other
  below. The flow is sub-fast magnetosonic in the lower right region;
  therefore no characteristics exist there and signals can propagate
  freely.}
\end{figure}

\begin{figure*}
  \resizebox{\hsize}{!}{\includegraphics{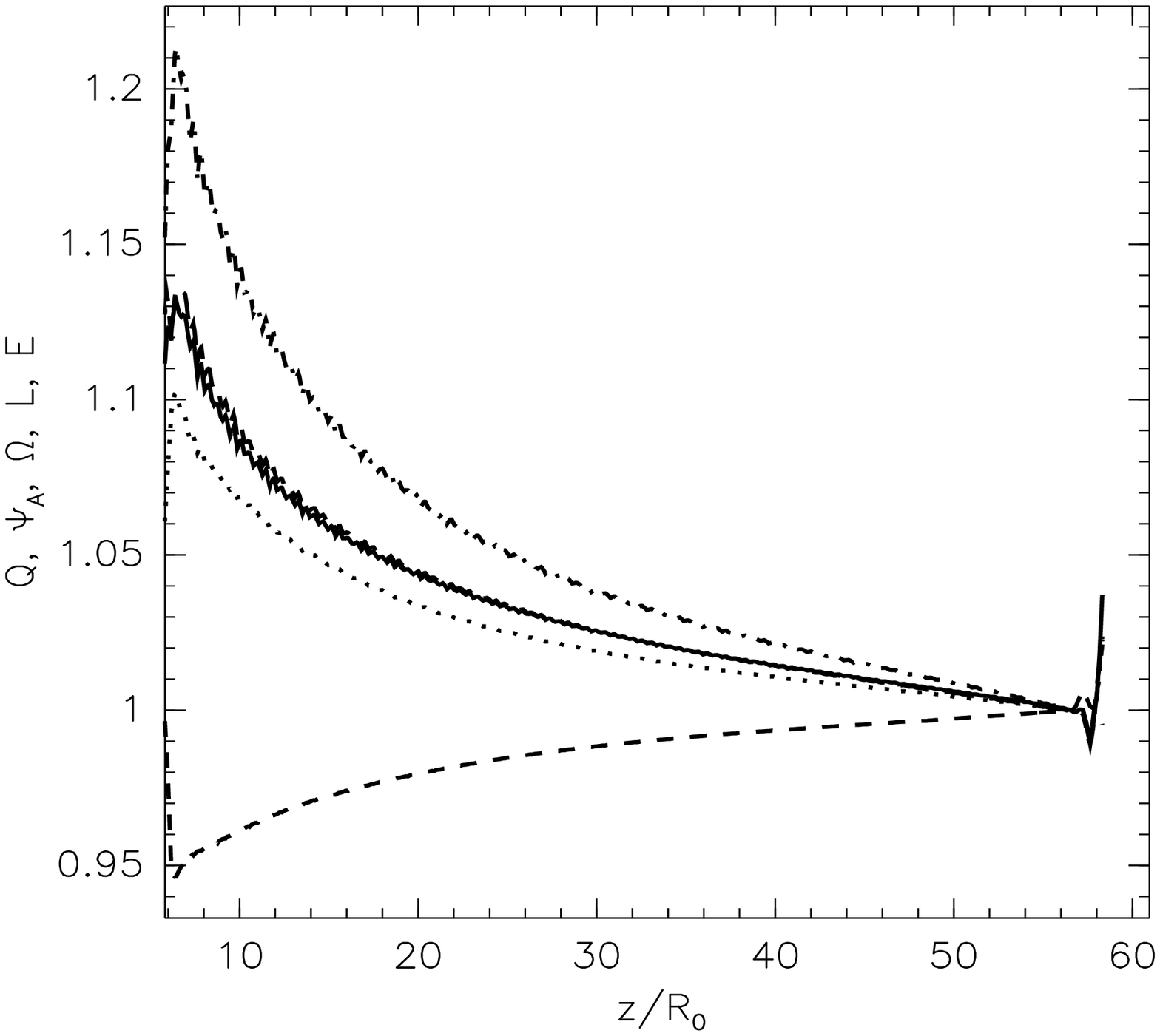}%
    \includegraphics{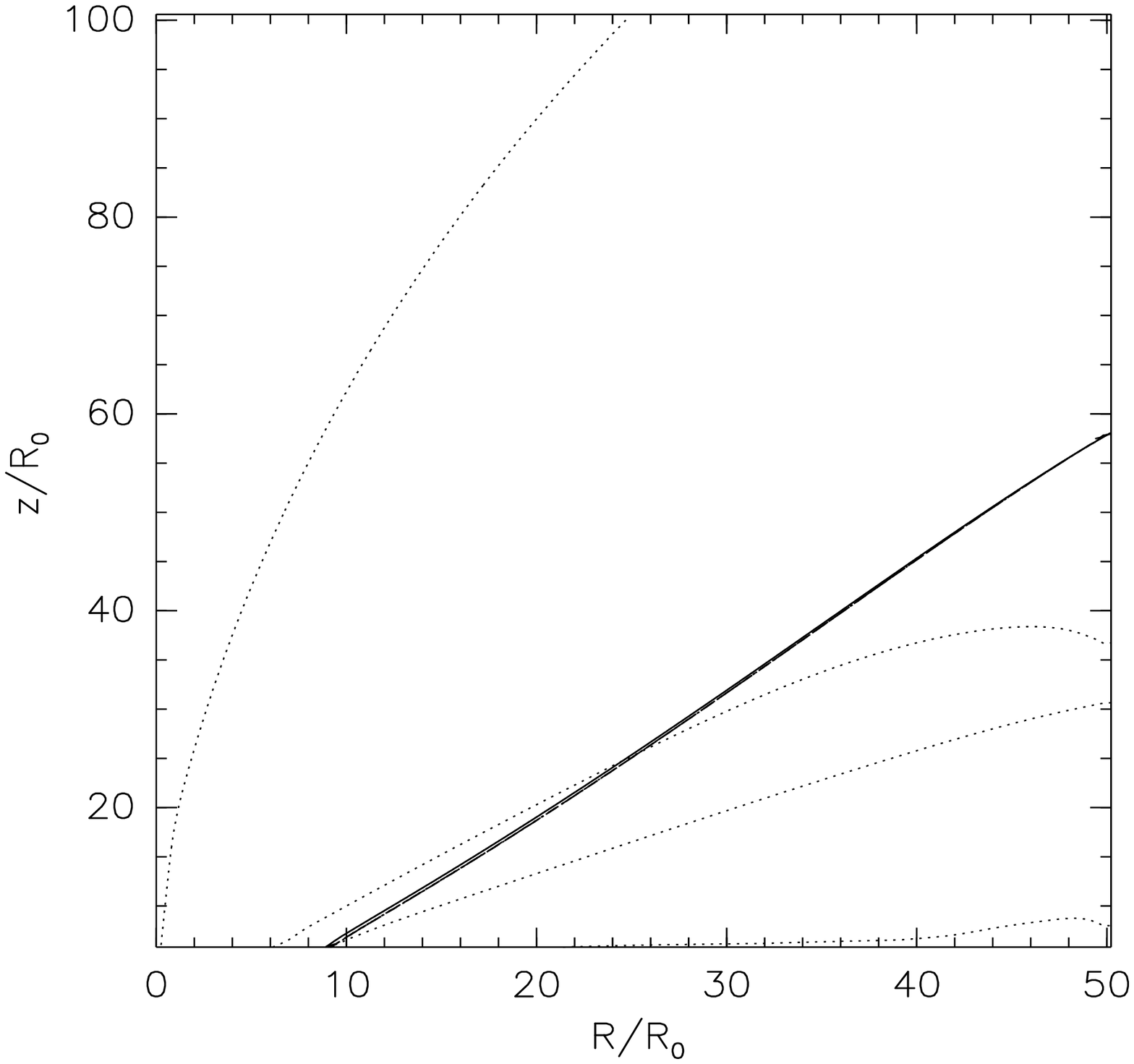}}
  \caption{ \label{fig:integrals} Evolution of the integrals of motion
  along a typical fieldline at intermediate latitudes for the final
  state. The left plot shows, from top to bottom, the
  integrals angular velocity $\Omega$ ({\em dot-dashed}),
  total energy $E$ ({\em long-dashed}), entropy $\entrop$ ({\em solid}), mass-
  to magnetic flux ratio $\kflux$ ({\em dotted}) and specific angular
  momentum $L$ ({\em short-dashed}) along the fieldline. The curves
  for entropy and total energy are almost identical. The values of
  each curve have been normalised to unity near the top boundary.  The
  right plot shows the position of the integral lines and the
  fieldline in the poloidal plane. The values chosen for the integrals
  are the same, as the ones for the normalisation in the left
  plot. The integral lines are almost indistinguishable. The dotted
  lines indicate the critical surfaces of the final model as in figure
  \ref{fig:MHD_surfaces}.}
\end{figure*}

In this section we discuss the results of a particular simulation
which shows the properties of a typical run. The model parameters are
chosen as $(x, \lambda^2, \mu, \kappa, \gamma) = (0.75, 136.9, 2.99,
2, 1.05)$. This particular model has been discussed in detail in
\citet{\VTST}. The simulation evolves for several Alfv\'enic crossing
times, such that the final state can be considered as relaxed.

\subsection{Overall relaxation}
The initial conditions are not a steady solution of the system of
equations under consideration. Therefore, the initial conditions will
relax toward a final steady state, if this exists, in accordance with
the boundary conditions. These do differ, as discussed in the previous
section, from the boundary conditions of the analytical model. So we
expect the final numerical state to be different from the analytical,
also. Furthermore, the analytical model is not valid near the rotation
axis. The numerical model includes the axis consistently, where the
initial conditions will likely deviate most from the final state.

As expected from the high MHD signal velocities, the inner regions of
the flow evolve very rapidly towards the final state.  MHD-waves
communicate changes of the inner flow to the outer regions and are
clearly visible as bends moving along the fieldlines. After a couple
of Alfv\'enic crossing times, the computational domain does not change
substantially anymore and the simulation is halted.

\subsection{Critical MHD surfaces and characteristics}
A classical MHD critical surface is the location of points where the
poloidal velocity $V\pol$ equals the phase speed of one of the MHD
waves propagating along the flow.  There are three such surfaces
corresponding to the three MHD waves, namely the slow magnetosonic,
Alfv\'en, and fast magnetosonic, with phase speeds $\vs$, $\vap$, and
$\vf$, respectively.  Figure \ref{fig:MHD_surfaces} shows the location
of the MHD critical surfaces for the initial and the final model. In
the initial state the slow magnetosonic surface is located at some
distance from the lower boundary; in the final state, it almost
disappears bellow the boundary.

The most important of the three MHD surfaces is the fast one, because
it is related to the causal connection between the asymptotic part of
the flow and the upstream (sub-fast) regime.  In the super-fast regime
signals can only propagate within a Mach cone around the direction of
the poloidal velocity vector. The opening angle of this cone is a
function involving the propagation speeds of the MHD waves and the
flow speed. The two sides of the cone coincide with the local
characteristic curves of the MHD equations. Since the flow speed as
well as the MHD wave speeds change from point to point the Mach cone's
opening angle is not constant and the final area to which a signal
propagates -- and which is causally connected with the origin of the
signal -- does not resemble a cone, in general, as seen in figure
\ref{fig:characteristics}.  This is similar to the light cone in
Penrose diagrams.

Due to the topology of the characteristics near the classical fast
surface, a Mach cone with its origin in the super-fast regime of the
flow may end on the fast surface, below which a signal may propagate
freely. Thus, a signal that originates in the super-fast regime may
affect the sub-fast regime.  This ceases to be the case downstream
from a limiting characteristic, or fast magnetosonic separatrix
surface (FMSS), or, modified-fast surface, for short
\citep{1994MNRAS.270..721B, 1996MNRAS.283..811T}. It can be shown that
the limiting characteristic coincides with the fast surface only in
the case where the poloidal velocity is normal to the latter.  The
underlying analytical model has a conical modified-fast surface
\citep{\VTST}.  The existence and location of this surface is very
sensitive to the topology of the magnetic field and thus is easily
destroyed when modifying the analytical solution to obtain the initial
numerical model and boundary conditions. Still the modified-fast
surface is present in the initial model near the axis, as seen in
figure \ref{fig:MHD_surfaces} as a line at constant angle. While the
classical MHD surfaces do not change much away from the boundary, the
modified-fast changes its location significantly.  As is shown in
figure \ref{fig:characteristics}, all characteristics converge to the
limiting one at both ends, i.e. near the origin and far away from the
base of the flow. In the self-similar solution this can be observed
only near the origin.

Since the numerical boundaries are not completely transparent to
magnetosonic waves, these will artificially influence the solution
near the edge of the computational box. This is evident, for example,
in the downwards bend of the critical surfaces near the outer $\R$
boundary, or possibly even the disappearance of the slow magnetosonic
surfaces.

\subsection{MHD integrals of motion} 
It can be shown, that steady, axisymmetric, ideal MHD flows
conserve several physical quantities along the fieldlines. This means
that these so called {\em integrals of motion} are a function of the
magnetic flux $\Amag$ only. The integrals are the entropy 
function $\entrop$, the mass-to-magnetic-flux ratio $\kflux$, the
total angular velocity $\Omega$, the total specific angular momentum
$L$ and the total energy-to-mass flux ratio $E$. These integrals are
given as
\begin{equation} 
  \entrop = P/\rho^\gamma,
\end{equation}
\begin{equation}
  \kflux = 4\pi \frac{\rho V\pol}{B\pol},
\end{equation}
\begin{equation} 
  \Omega = \frac{1}{\R} \left(V_\phi - \frac{B_\phi V\pol}{B\pol}\right),
\end{equation}
\begin{equation}
  L = \R \left(V_\phi - \frac{4\pi}{\mu_0} \frac{B_\phi}{\kflux}\right),
\end{equation}
\begin{equation} \label{eq:int_E}
  E = \frac{V^2}{2} + \frac{\gamma}{\gamma-1} \frac{P}{\rho} + \grav
  - \Omega \R \frac{4\pi}{\mu_0} \frac{B_\phi}{\kflux}.
\end{equation}

Due to the modifications applied to the magnetic field, the
fieldlines do not coincide with the integral lines in the initial
conditions. During the relaxation process the structure of the flow
changes and a realignment can be observed.
The realignment is not perfect as seen in the left panel of figure
\ref{fig:integrals}, where the values of the integrals are plotted
along a typical fieldline at intermediate latitudes. The integrals
vary along the fieldline, especially near the base, where they rise
(or fall) steeply, before approaching their final value
asymptotically. Fieldlines further out show less variation than
fieldlines further in.

This means, that the surfaces of the constant values of the integrals
are not exactly parallel to the surfaces of constant magnetic flux,
and further, that even some of the integrals are not parallel to each
other. In fact, only entropy and total energy are perfectly
aligned.

There are several possibilities to explain this discrepancy. First,
the calculation of the magnetic flux and its iso-surfaces is not
accurate enough. This does certainly play a role, since the magnetic
field is only interpolated linearly in the calculation of the magnetic
flux.  Another interpolation of low order is made to calculate the
contours.  So interpolation errors sum up. To check, the magnetic
fieldlines have been calculated with a different method as the
streamlines of a mass-less particle flowing through a vector field
given by the magnetic field. This shows similar behaviour of the
integrals. On the other hand, the relative change of the integrals
along the fieldline is only up to 20\%, while the integrals change by
several orders of magnitude across the fieldlines. So the error in the
calculation is actually low, but should still be accounted for. This
is clearly illustrated by the right plot in figure
\ref{fig:integrals}, which shows the location of all integral lines
and the fieldline in the poloidal plane. They are virtually
indistinguishable. Fieldlines near to the axis generally show higher
miss-alignment than fieldlines further out. The integral lines always
match each other almost perfectly.

The next source of potential inaccuracies is the nature of the
numerical code NIRVANA. This is a non-conservative, grid-based MHD
code. As such it does not guarantee the conservation of energy- or
momentum-flux across grid-boundaries down to machine precision
level, and leads to numerical dissipation. The degree of numerical
dissipation can be estimated by convergence studies -- as done by
\citet{2001A&A...380..789K} in the context of jet propagation -- which
show, that NIRVANA conserves energy and momentum to a reasonable
level. Our own calculations at different resolutions confirmed
this. Further, very high numerical resistivity would rather lead to an
increase of the miss-alignment of the integral lines along the flow
instead of an realignment as observed in our simulations.

Most probable is though, that the miss-alignment near the base is due
to the mathematically overspecified boundary conditions
\citep[e.g.][]{2003ApJ...595..631K}.

\subsection{Acceleration and collimation} 
\begin{figure}
 \resizebox{\hsize}{!}{\includegraphics{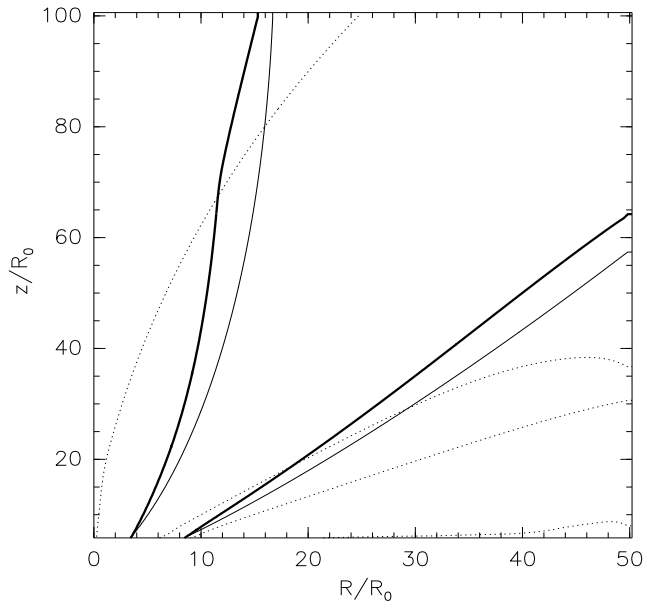}}
  \caption{ \label{fig:Ecurves} 
  Surfaces of constant conserved total energy $E$ in the poloidal
  plane for two different energy values. The thin lines show the
  initial state, while the final state is shown in heavy lines. The dotted
  lines indicate the critical surfaces as in the previous
  figures. During the relaxation phase the integral lines and
  fieldlines reallign. In the final state integral lines are almost
  perfectly parallel to the magnetic fieldlines.}
\end{figure}

\begin{figure}
  \resizebox{\hsize}{!}{\includegraphics{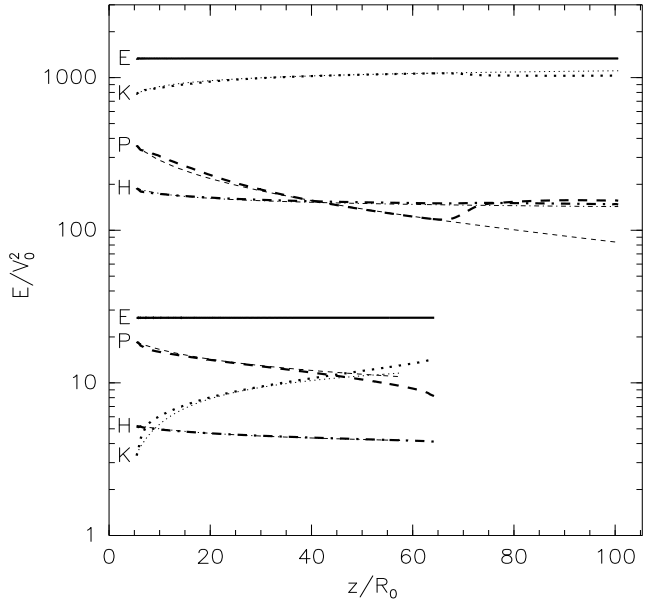}}
  \caption{ \label{fig:Eterms} 
  Evolution of the individual contributions to the total energy $E$ on
  the integral lines shown in figure \ref{fig:Ecurves}. Again, thin
  lines indicate the initial state, while heavy lines the final state.
  The two clearly separated sets of lines belong, from top to bottom,
  to the inner and the outer integral line in figure
  \ref{fig:Ecurves}.  The total energy is shown as solid (also, label
  E) line, the other line types indicate the different contributions
  as kinetic energy ({\em dotted/K}), enthalpy ({\em dot-dashed/H})
  and Poynting flux ({\em dashed/P}). The gravitational energy is not
  shown, since it is orders of magnitude lower.
  }
\end{figure}

\begin{figure}
  \resizebox{\hsize}{!}{\includegraphics{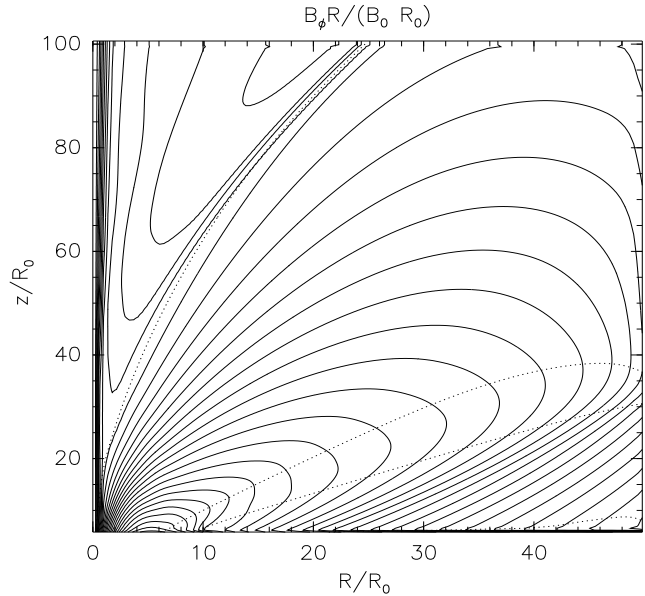}}
  \caption{ \label{fig:currents} Logarithmically spaced contours of
  the enclosed poloidal current $B_\phi \R$ of the final model as
  solid lines. The dotted lines indicate the MHD critical surfaces as
  in the previous plots. The absolute current density is highest near
  the lower left corner. Beyond the modified-fast surface, the current
  shows a local maximum, before decreasing again sharply towards the
  axis.}
\end{figure}
\begin{figure}
  \resizebox{\hsize}{!}{\includegraphics{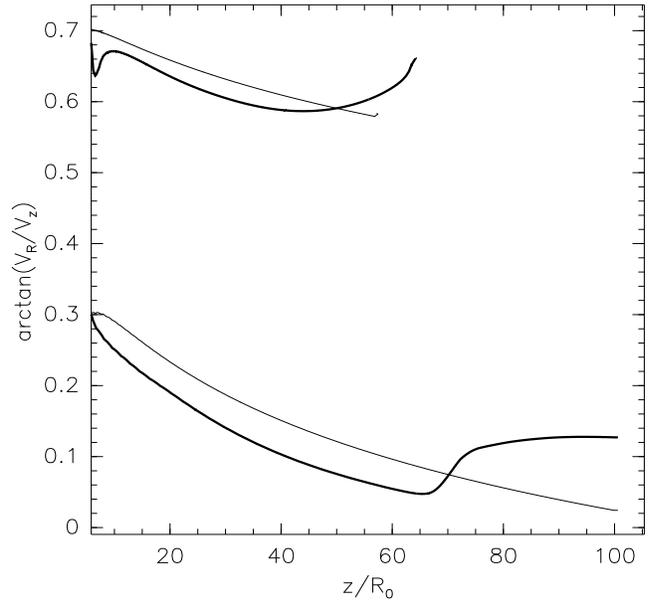}}
  \caption{ \label{fig:angle} 
    Angle between the flowline and the rotation axis $\delta = \arctan
    V_\R/V_\Z$ along the same fieldlines as shown in figure
    \ref{fig:Ecurves}. The thin lines show the initial state, while
    the final state is shown as heavy lines. The upper set of curves
    corresponds to the outer fieldline, and the lower to the inner.}
\end{figure}

The contributions to the total energy $E$ are, from left to
right in equation \Ref{eq:int_E}, the kinetic energy, the enthalpy,
the gravitational potential and the Poynting flux. Strictly speaking,
all these quantities are energy-to-mass flux ratios, but
we will henceforth drop this cumbersome notation and refer to them in
this simplified form. Since the total energy is conserved along a
fieldline, acceleration of the flow implies an increase of kinetic
energy along the fieldline at the expense of other forms of energy.
At the base of the fieldline, in the sub-slow region, the kinetic
energy will be lower than the Poynting flux or enthalpy, while at the
top, where the flow is super-fast magnetosonic, the total energy is
dominated completely by the kinetic energy. Thus, along the flow
different sources of energy will be transformed into kinetic
energy. The dominant conversion channel depends on the position on the
fieldline in relation to the critical surfaces. In principle each
fieldline passes through all phases. But since the spatial resolution
and extent of our simulation box is limitied, we will have to show the
different conversion channels on different fieldlines.

Figure \ref{fig:Ecurves} shows the integral lines of the total energy
for two different values of $E$ for the initial and the final
state. We assume for a moment, that the integral line and the
fieldline are identical -- which is actually close to reality as
previously discussed -- and limit our interest to the integral lines
only. This makes the comparison of initial to final model easier,
since the value of the total energy does not change, but we rather
jumped from one fieldline to the other.  Figure \ref{fig:Eterms} shows
the different contributions to the total energy integral
(eq. \ref{eq:int_E}) along these fieldlines.  At the base of the outer
fieldline the Poynting flux dominates over the enthalpy and kinetic
energy, the lowest contribution, apart from gravitational energy which
is not shown. Along the fieldline the flow gains kinetic energy at the
expense of Poynting flux and, to a lesser degree, enthalpy. At the
outer boundary of the computational domain, the flow is already
kinetically dominated due to acceleration through the Lorentz
force. The initial model and the final state look very similar. In the
final state the acceleration is slightly more efficient, especially
near the base.

Two different mechanisms are responsible for the acceleration. Bellow
the Alfv\'enic surface, the {\em magneto-centrifugal mechanism} taps
into the rotational energy of the magnetic fieldlines, which are {\em
stirred} at their base by the underlying accretion flow, and propels
the plasma outwards and upwards through centrifugal forces like a {\em
bead on the wire} \citep{\BP,1992ApJ...394..117P}. Beyond the
Alfv\'enic surface this mechanism is inefficient. The flow is
accelerated instead directly by the gradient of the toroidal magnetic
pressure which builds up because of the inertia of the fluid. The
energy stored in the tightly bound-up magnetic field loops acts
similarly to an {\em uncoiling spring} \citep{1985PASJ...37..515U} and
is transfered into kinetic energy of the plasma parcels.

Near the modified-fast surface the flow properties change
character. The upper set of curves in figure \ref{fig:Eterms} shows
the energy contributions on a fieldline which crosses the
modified-fast surface at a height of approximately $70\, \R_0$. This
fieldline is already kinetically dominated at the lower boundary of
the computational domain. The acceleration continues until the flow
crosses the FMSS. There, kinetic energy is converted back into
Poynting flux and the flow slightly decelerates.  The
equipartion between enthalpy and Poynting flux is coincident and not
typical. This change of character of the flow is not seen in the
initial state, neither for this particular fieldline, which did not
cross the FMSS, nor for fieldlines further in, which do cross the
FMSS. The accelaration is not resumed beyond the modified-fast
surface. 

The collimation process is best understood in terms of poloidal
current lines as shown in figure \ref{fig:currents} for the final
state. The azimuthal magnetic field component $B_\phi$
points out of the plane, and the poloidal current $\vec j\pol$ flows
in the counter-clockwise sense along the indicated current
lines. Let's focus first on the region below the modified-fast
surface.  In this region all the poloidal current lines close near the
equator outside of the simulation box.  The poloidal component of the
Lorentz force, i.e. $\vec j\pol \times \vec B_\phi + \vec j_\phi
\times \vec B\pol$ points inwards on the ``left'' side of the current
loops and the flow is collimated.  We find decollimation on the
``right'' side of the currents loops, where the first term points
outward. However, this region is also affected by the outer boundary
and the decollimation may be, at least partly, a numerical artifact.

Beyond the modified-fast surface the enclosed poloidal current
$B_\phi\R$ takes a local maximum. Near the FMSS, the first term of the
poloidal Lorentz force points outward, while the second term $\vec
j_\phi \times \vec B\pol$ is negligible already beyond the Alfv\'enic
surface. The flow decollimates (and decellerates) while crossing the
modified-fast surface. After crossing the local maximum in the
azimuthal current distribution, i.e. on the falling side of poloidal
loops, the collimation is resumed, since both components point inward
again. However, the efficiency of collimation lowers continously,
since the flowline encounter decreasing absolute values of poloidal
current.

A different way to illustrate the collimation is to plot the angle
between the poloidal flow lines and the rotation axis $\arctan
V_\R/V_\Z$ as shown in figure \ref{fig:angle}. Again, we show values
along the same fieldlines as in figure \ref{fig:Ecurves}. In the
initial state, and for that matter in the analytical solution, the
opening angle steadily decreases along the flow. In the final state,
the opening angle increases strongly as the flow crosses the
modified-fast surface. Inspection of fieldlines further in, reveals
that the collimation is resumed but with low efficiency, as noted
before. The opening angle of the outer fieldline increases near the
outer boundary. This is, as discussed, most probably an artifact of
the outer boundary.

\begin{figure}
  \resizebox{\hsize}{!}{\includegraphics{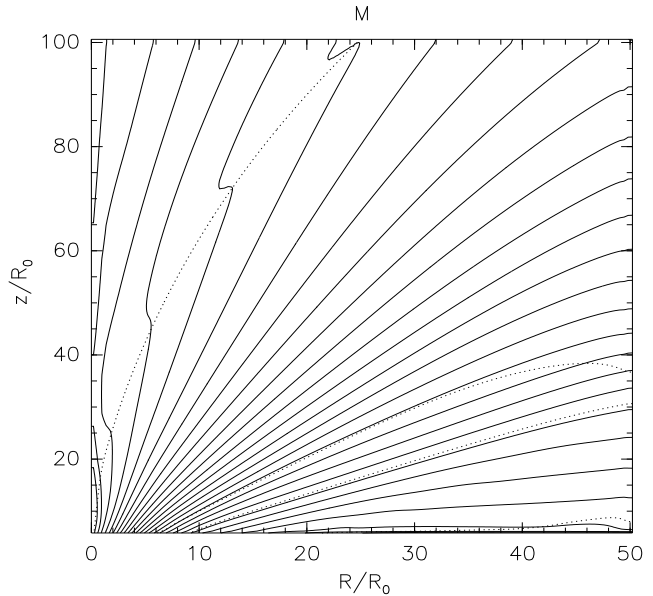}}
  \caption{ \label{fig:M} Logarithmically spaced contours of M. The
    dotted lines indicate the MHD critical surfaces as in the previous
    plots.}
\end{figure}

Finally, figure \ref{fig:M} shows the functions M of
the analytical solution, calculated from the final numerical state as
$$M_\num = \left(\frac{\mu_0\rho V\pol^2}{B\pol^2}\right)^{1/2}.$$
Note that some minor differences from the conical shape of the
analytical function exist only near the outer boundary and beyond the
modified-fast surface.

\section{Discussion}

\subsection{Robustness of the model}
We found that the results of the numerical simulation presented in
this paper are robust. The conclusions do not hold exclusively for the
presented numerical model, i.e. initial conditions and boundary
conditions. For example, we explicitely performed simulations with
very different initial conditions for the computational domain
including 1) rotation of the disk starts at $t=0$, i.e. $B_\phi = 0,
V_\phi = 0$ initially; 2) vertical poloidal magnetic field, i.e. $\vec
B\pol = B_\Z \eu_\Z$; 3) vertically non-stratified medium, i.e.
$(\rho, P) = f(\R, \Z=0)$; and combinations thereof. The results are
qualitatively similar. Small quantitative differences arise through
the influence of the downstream boundaries. That means, that the
conclusions are independent of the initial conditions as desired. The
final state is indeed given only by the propagation of the imposed
boundary conditions into the computational domain.

Further, the properties are robust to same changes in the boundary
conditions. For example, in this work we imposed the magnetic field
component $B_\R$ from the analytical solution on the lower $\Z$
boundary, and extrapolated the other from the computational domain
under the $\Div \vec B$-free constraint. We also performed
simulations, where we instead imposed the magnetic flux function $A$
from the analytical solution, with the same qualitative results. We
also relaxed the number of boundary conditions to impose (see
discussion further down) and obtained similar results. Albeit, we
could not reduce the number of boundary conditions to the desired one
for numerical reasons specific to this particular profile of the
azimuthal vector components on the lower boundary. The aim of this
work was, to stay as closely as possible to the analytical model
without introducing more differences in the models than necessary.

As discussed in the previous paragraphs, the numerical results are
unaffected by certain changes to the initial conditions and the
boundary conditions imposed {\em upstream}. In the following we will
offer some remarks on the influence of the boundary conditions {\em
downstream}. For numerical reasons we have to provide values for the
physical quantities on the downstream boundaries as well. Ideally,
these boundaries should be transparent and should not influence the
computational domain artificially, since they exist only as numerical
boundaries beyond which the numerical code does not solve the
equations; they are not related to real physical boundaries in any
way.

One way to judge the possible impact of numerical boundaries is to
study the behaviour of MHD waves propagating through the
boundary. Again, ideally, the waves should propagate unhindered
and no part should be reflected. Although the numerical code we use
does not take special measures to inhibit the reflection of waves,
numerical experiments show, that in a super-fast magnetosonic
situation, no magnetosonic waves are reflected at all. In the sub-slow
regime, all waves may be reflected partially, but the amplitude of the
reflected wave is in general very low compared to the incident. We
note further, that the reflected waves propagate, if at all, only in
accordance with their respective Mach cones. No artificial propagation
of waves along the boundary is observed. Due to the topology of the
Mach cone in our model (see figure \ref{fig:characteristics}), this
means, that no wave traversing the upper $\Z$ boundary can propagate
back into the computational domain and influence it. The same is true
for the upper $\R$ boundary beyond the intersection with the classical
fast surface. In summary, only the upper $\R$ boundary below the
classical fast surface can affect the computational domain.

As a side note we remark, that due to the shape of the Mach cone, most
of the classical super-fast region cannot influence the sub-slow
region, also, since the Mach cone intersects the classical fast
surface only outside of the computational box. In fact, only points
below the characteristic, which originates at the intersection of the
upper $\R$ boundary and the classical fast surface have Mach cones
which reach into the sub-fast region within the computational
domain. The Mach cones originating in points above this characteristic
do not reach into the sub-fast region before encountering the outer
boundaries.

Another aspect to consider is the number of boundary conditions to
impose at the upstream boundary.
Following \citet{1997A&A...323..634B}, the number of boundary conditions is
to be reduced by one for each outgoing wave propagating perpendicular
to the boundary.
In the present work, we over-specified the lower $\Z$ boundary,
i.e. we imposed more conditions than mathematically allowed, even if
one does not take into account the ambiguity with regard to the nature
of the critical surfaces. We chose to do so mainly for technical
reasons specific to this particular model, as discussed elsewhere in
this paper. The only artifact that we may attribute to the
over-specification of the launching-surface, is the steep increase
(decrease) of the integrals near the base of the fieldlines, see
figure \ref{fig:integrals}, where the flow tries to match the integral
lines given by the conditions at the critical surfaces to the
conditions imposed artificially on the boundary. See also
\citet{2003ApJ...595..631K} on the risk of creating discontinuities by
over-specification of the boundary conditions.  We note however, that
the implementation of a relaxed number of boundary conditions has been
tested successfully with other models, and will be presented and used
in future work.

\subsection{Comparison to related studies}
In general, the analytical solution \citep{\VTST} and our numerical
solution are quite similar, as expected from the way the latter is
constructed. Differences arise only where the self-similarity
assumption is violated. We explicitly included the rotation axis into
our analysis. There the flow properties are mainly given by the
symmetry conditions on the axis, which are not compatible with radial
self-similarity, in general, due to their diverging behaviour. In our
numerical model all physical quantities match their finite value on
the axis smoothly. 
The numerical model solves the problem of the termination of most
radially self-similar analytical solutions at finite distance along
the flow.
In particular, the $\R$-component of the flow velocity decreases
continously and eventually becomes negative. At larger distances, even
the $\Z$-component changes sign as well.
So, technically the solution does not extent to infinity, but
terminates at finite distance along the flow. This behaviour is not
present in the numerical model. Neither do the fieldlines bend
inwards, nor would this automatically mean, that the flow terminates
at the axis, since the azimuthal magnetic field component $B_\phi$
vanishes at the axis and the Lorentz force does likewise.

The only region where substantial difference arise is near the
modified-fast surface. At this surface the flow just looses causal
contact with the base for the analytical model, but no locally
evaluated quantity behaves differently than before crossing this
surface. This is not the case in the numerical model. Again, the flow
looses causal contact with the base, but unlike the analytical
solution, in some way it also looses its memory of the flow
properties. In fact, all relevant physical quantities change at the
critical surface and find a new equilibrium value. On the other hand,
there is no shock present either. In principle a terminated solution
can connect to infinity by going through a shock beyond the critical
surface. This does not seem to be the case here, since, although all
physical quantities change noticeably, the involved gradients are
moderate. The behaviour of the flow near the modified-fast surface is
closely related to the presence of a local maximum in the enclosed
poloidal current distribution further down the flow. This temporarily
reverses the direction of the dominant part of the Lorentz force $\vec
j\pol \times \vec B_\phi$ and therefore decelerates and decollimates
the flow noticeably. It remains to be checked if this is a general
behaviour of solutions crossing the modified-fast surface.

One qualitative difference between our model and \citet{\KLBi} the
shape of fast magnetosonic surface. In our case the flow is
classically super-fast magnetosonic everywhere near the axis, in fact
even on part of the launching surface. In contrast, the axial region
in \citet{\KLBi, \KLBii} is occupied by a cold, light, sub-fast
magnetosonic flow injected at the lower boundary, apparently for
numerical reasons. Further, our model crosses not only the
classical, but also the modified-fast surface. The structure of the
magnetic field is very similar, only after crossing the FMSS, does the
magnetic field tend to open up. This is neither described elsewhere,
nor is it understood so far and will be the subject of further
investigations.

By extrapolating the experience we obtained in this study, the shape
of the Mach cones originating from the downstream boundaries may
influence the solution in the computational domain, since these
boundaries may still be in causal contact with the base of the flow.
This might explain the sensitivity of \citeauthor{\KLBi} and others model to
changes of the computational box size -- a problem that we did not
encounter \citep[see for example][for a
discussion]{1995ApJ...439L..39U}.
A direct consequence of the shape of the super-fast surface is the
fact, that our models reach higher fast-Mach numbers $M_{\rm f}$ --
typically $\sim 5$ at the upper boundary and in excess of $10$ near
the axis -- and therefore the flow is more efficiently accelerated. On
the other hand a high poloidal magnetic field strength at the axis,
i.e. low Mach numbers, seems to stabilise the flow against
non-axisymmetric Kelvin-Helmholtz instabilities as discussed by
\citet{2003ApJ...582..292O} and more generally by
\citet{1992A&A...256..354A}, \citet{2005A&A...430....9B} and
references therein. The mass-loading is however given by the
conditions on the launching surface as \citet{\KLBii} showed very
convincingly.


Recently, \citet{2004ApJ...601...90C} studied resistive MHD
accretion flows threaded by a vertical magnetic field and achieved
near-steady state outflows along the axis. Their analysis includes the
accretion flow dynamics and does not rely on assumed ad-hoc boundary
conditions for the disk at a launching surface. Similar to our model
and unlike \citet{\KLBi}, the axial region is not taken by an
artificial cold axial jet but rather self-consistently by a hot
outflow. This allows the outflow to cross the classical fast surface
even on the axis. However, it is not clear whether it also crosses the
modified-fast surface and what the structure of the characteristics
is. Further, the enthalpy flux seems to play a important role at the
base of the flow, while our model is Poynting flux dominated at the
base and shows very high accelerations efficiency along the outflow.

\subsection{Conclusions and future plans}

In this paper, we verified the analytical model presented by
\citet{\VTST} with numerical methods. Further, we extended the
analysis into the domain where the analytical model breaks down due to
the self-similarity assumption, i.e. we included the rotation axis in
a self-consistent way. Unlike the analytical solution, our numerical
solution is a global one in the sense, that it is not terminated but
extends -- given enough computational resources and infinite
computational box size -- to infinity. We confirmed the properties of
the analytical solution in its domain of availability, in particular
the existence of the modified-fast surface or fast magnetosonic
separatrix surface. The numerical solution shows different properties
near and beyond the modified-fast surface, where the flow changes
character. This change of character manifests itself in the appearance
of a further local maximum in the enclosed poloidal current beyond the
modified-fast surface, which is responsible for a temporary
deceleration and decollimation of the flow.
 
This work is intended to be the first in a series of papers. We will
subsequently address some of the shortcomings of this work by
lovering the launching surface down to the equator and applying only
the neccesary number of boundary conditions. 
In this way it will be possible to explore a wide range of scenarios
for the jet-launching surface.


\section*{Acknowledgements}
Part of this work was supported by the European Community's
Research Training Network RTN ENIGMA under contract HPRN-CY-2002-00231.

\bibliographystyle{mn2e}
\bibliography{papers} 


\label{lastpage}
\end{document}